\documentclass[a4paper,11pt]{article}
\usepackage{jheppub} % for details on the use of the package, please see the JINST-author-manual
\usepackage{lineno}
\usepackage{physics}
\usepackage{amssymb}
\usepackage{amsmath}
\usepackage{mathtools}
\usepackage{tikz}
\usetikzlibrary{chains}

\tikzset{node distance=2em, ch/.style={circle,draw,on chain,inner sep=2pt},chj/.style={ch,join},every path/.style={shorten >=4pt,shorten <=4pt},line width=1pt,baseline=-1ex}

\let\dlabel=\alabel

\newcommand{\dnode}[2][chj]{%
\node[#1,label={below:\dlabel{#2}}] {};
}

\newcommand{\dnodebr}[1]{%
\node[chj,label={below right:\dlabel{#1}}] {};
}

\newcommand{\dnoder}[2][chj]{%
\node[#1,label={right:\dlabel{#2}}] {};
}

\newcommand{\dydots}{%
\node[chj,draw=none,inner sep=1pt] {\dots};
}

\arxivnumber{2304.11830} % if you have one

\author{Chao Ju}
\affiliation{Berkeley Center for Theoretical Physics and Department of Physics\\
University of California, Berkeley, CA 94720, U.S.A.}

\emailAdd{cju19@berkeley.edu}

\abstract{The Hilbert space of level $q$ Chern-Simons theory of gauge group $G$ of the ADE type quantized on $T^2$ can be represented by points that lie on the weight lattice of the Lie algebra $\mathfrak{g}$ up to some discrete identifications. Of special significance are the points that also lie on the root lattice. The generating functions that count the number of such points are quasi-periodic Ehrhart polynomials which coincide with the generating functions of $SU(q)$ representation of the ADE subgroups of $SU(2)$ given by the McKay correspondence. This coincidence has roots in a string/M theory construction where D3(M5)-branes are put along an ADE singularity. Finally, a new perspective on the McKay correspondence that involves the inverse of the Cartan matrices is proposed.}

\title{Chern-Simons Theory, Ehrhart Polynomials, and Representation Theory}

\begin{document}
\maketitle
\flushbottom

\section{Introduction}
\label{sec:intro}
The 3-dimensional Chern-Simons theory is a remarkably rich tool that has applications in both mathematics and physics. For example, Witten showed that Chern-Simons theory is connected to the Jones polynomial and knot invariants~\cite{wittenjones}. It is also applied in the study of S-duality~\cite{ganorandhong,gaiottoandwitten}, a subject related to the geometric Langlands program~\cite{wittenkapustin}. In terms of the more tangible physics applications, Chern-Simons theory has been used extensively to study physics on 2-dimensional surfaces. For example, it is used to endow particles in 2+1 dimensions with fractional statistics~\cite{wilczek,wilczekandzee}, so that upon exchanging two identical particles the wave function of the two particles can end up with a phase different from $\pm 1$. In addition, Laughlin explained fractional quantum hall effect by applying Chern-Simons theory\footnote{For a review of Chern-Simons theory and fractional quantum hall effect, see~\cite{zee} and the references therein.}~\cite{laughlin}.

The fact that Chern-Simons theory is important for 2+1 dimensional physics is no accident. The Chern-Simons action 
\begin{equation}
\label{eq:a1}
S = \frac{q}{4\pi} \int \text{tr} \left(A\wedge dA + \frac{2}{3} A\wedge A\wedge A\right)
\end{equation}
is the unique relativistically invariant action of the gauge field $A$ that has only one derivative acting on $A$. In this paper, we take the convention that $A$ is $\mathfrak{g}$ valued 1-form and the trace is taken in the representation such that $q$ is a quantized positive integer. We take $\mathfrak{g}$ to be simply laced, namely $\mathfrak{su}(N)$, $\mathfrak{so}(2N)$, $\mathfrak{e}_6$, $\mathfrak{e}_7$, and $\mathfrak{e}_8$. The global property of the gauge group will not concern us in this paper. The geometry we are interested in is $T^2 \times \mathbb{R}$ where $\mathbb{R}$ is the time direction. Upon quantizing the Chern-Simons theory in this geometry, one gets a discrete set of states that can be identified with points in the set~\cite{cs}
\begin{equation}
\label{eq:a2}
\frac{\Lambda_w}{W\ltimes q\Lambda_r}
\end{equation}
where $\Lambda_w$, $\Lambda_r$ are the weight and the root lattice of $\mathfrak{g}$, and $W$ is the Weyl group. \par
In this paper, we study the counting of a special class of states that lie in the set
\begin{equation}
\label{eq:a3}
\frac{\Lambda_w}{W\ltimes q\Lambda_r} \cap \Lambda_r
\end{equation}
where the intersection with $\Lambda_r$ picks out the states in the set~\eqref{eq:a2} that are also roots.

We will show that the counting of such states leads to a curious connection to Ehrhart polynomials, McKay correspondence, and representation theory. Ehrhart polynomials were first constructed to count lattice points in rational polytopes~\cite{ehrhart}, a problem that is in general NP-hard~\cite{deloera} to solve by computer. Ehrhart polynomials have been found to connect different areas of mathematics such as number theory, geometry, and topology\footnote{For an extensive introduction, see the book~\cite{beckandrobins}}. Since closed-form Ehrhart polynomials are rare and are only for very special geometries, our computation in this paper will add more examples to the known collection of Ehrhart polynomials.

We will also show that the Ehrhart polynomials we obtain from the geometric point of view can also be obtained from a representation theory point of view. In fact, the latter comes from a dual formulation of the problem. The two approaches are connected by the McKay correspondence~\cite{mckay}, which gives a ADE Dynkin diagram classification of discrete subgroups of $SU(2)$ and their irreducible representations. The physics behind this dual formulation of the problem comes from a holographic system of D-branes on ADE singularity. More precisely, there is a duality between the ground state Hilbert space $SU(q)$ $\mathcal{N}=4$ Supersymmetric Yang-Mills theory on $S^3/\Gamma$ and a certain subspace~\eqref{eq:a3} of the Hilbert space of level $q$ Chern-Simons theory on $T^2$ with gauge algebra $\mathfrak{g}(\Gamma)$ given by the McKay correspondence. 

This paper is organized as follows. Because our construction has to do with D-branes on ADE singularity, in section~\ref{sec:review} we will review this subject and point out the difference between our construction and the literature. In section~\ref{sec:2}, we review the quantization of Chern-Simons theory on $T^2$ and formulate the Hilbert space geometrically in terms of certain points on the Lie algebra lattice. To illustrate the rather abstract notation, we give an example of states of $\mathfrak{g}=\mathfrak{su}(3)$. In section~\ref{sec:3}, we pose the problem of counting the number of the special class of states in the set defined in~\eqref{eq:a2}. We shall find that the problem is the same as counting lattice points in rational polytopes, and that the generating function for counting the states is exactly the corresponding Ehrhart polynomial. In section~\ref{sec:4}, we compute the explicit form of the Ehrhart polynomial for a specific case by using the $\Omega$ operator introduced by MacMahon~\cite{macmahon}. The general case is solved by reverse-engineering some representation theory formulas in section~\ref{sec:5}. We will see that the Ehrhart polynomial that counts the special states at level $q$ with gauge algebra $\mathfrak{g}$ is the same as the generating function for the $SU(q)$ representation of the ADE subgroup given by the McKay correspondence. In section~\ref{sec:6}, we extend our result to the D-series where the gauge algebra of Chern-Simons theory is $\mathfrak{so}(2(N+2))$, $N\geq 1$. In section~\ref{sec:7}, we discuss some curious representation theory properties implied by the inverse of Cartan matrices modulo 1 for ADE Lie algebras by focusing on the exceptional series $\mathfrak{e}_6$, $\mathfrak{e}_7$, and $\mathfrak{e}_8$. In section~\ref{sec:8}, we derive the string/M theory origin of the proposed duality.

The reader who is only interested in the physical aspect of this work can skip the second half of section~\ref{sec:3} and all of section~\ref{sec:4}.

\section{String Theory and D-branes on Orbifold}
\label{sec:review}
There is a long history of the study of string theory and D-branes on orbifold singularity. Since the particular orbifold singularity that will concern us in this paper is the ADE orbifold singularity, we will in this section do a quick review of some past research on ADE singularity and string theory. We will discuss how this work is different from past research and motivate the Chern-Simons/SYM duality mentioned in the introductory section~\ref{sec:intro}.

There are roughly speaking three uses of putting string theory or D-branes on ADE singularity. The first is to obtain a more realistic compactification of string theory down to lower dimensions. The second is to use string theory to probe the topology and geometry of the ADE singularity. The third is to use string theory construction to obtain new classes of quantum field theory.

One way to obtain ADE singulairty is the ``blow-down'' of K3 surface~\cite{stringorb}, the only nontrivial Calabi-Yau manifold in four (real) dimensions~\cite{aspinwall}. String theory on K3 surface leads to more realistic models of string compactification because the $SU(2)$ holonomy of K3 surface helps break half of the supersymmetry. It also leads to more examples of string dualities~\cite{aspinwall}. In fact, as we will see in section~\ref{sec:8}, the duality considered in this paper is derived using the duality between heterotic string on $T^4$ and IIA string on K3. 

The second use partially overlaps with the first use because the low energy spectra of string theory on K3 sheds light on the topological invariants of the K3 surface~\cite{stringorb,aspinwall}. To probe the geometry and not just the topology of K3 surface, D-brane technologies must be used~\cite{johnsonandmyers}. In~\cite{johnsonandmyers}, the metric of some ALE (asymptotically locally Euclidean) space was computed by probing the space using D1 branes \textit{transverse} to the space. The fact that the moduli space of vacua for the low energy theory coincides with the ALE space suggests that the metric of the underlying space coincides with the metric that enters the kinetic term of the low energy theory. 

The third use dates back to~\cite{stringcomments}, where it was mentioned that M-theory on ADE singulairty $\Gamma$ leads to a seven dimensional super Yang Mills theory transverse to the 4-dimensional ADE singularity with the gauge algebra $\mathfrak{g}(\Gamma)$ McKay-dual to the ADE singularity. In~\cite{douglas}~\cite{gomis}, quiver gauge theories are constructed by putting D3 branes \textit{transverse} to the orbifold singularity. In the language of AdS-CFT correspondence, in this construction, the singularity mods out the $S^5$ part of the bulk geometry and keeps the $AdS_5$ part untouched. 

The difference between this work and the previous constructions is that here, the ADE singularity is longitudinal to the D3 brane worldvolume. In other words, the bulk geometry is $AdS_5/\Gamma \times S^5$. As will be derived in section~\ref{sec:8}, the long distance limit of this holographic system will involve a Chern-Simons theory in the AdS bulk direction. Constructions similar to this work can be found in~\cite{dijkgraaf}\cite{vafawitten}. Differences between our construction and those in~\cite{dijkgraaf}\cite{vafawitten} will be explained in section~\ref{sec:8}.

\section{Quantization of Chern-Simons Theory on $T^2$ and the States}
\label{sec:2}
The Chern-Simons theory with gauge algebra $\mathfrak{g}$ on $T^2$ can be quantized using the standard quantization procedure by choosing a gauge $A_0=0$ and imposing the constraint $\delta S/\delta A_0 =0$, where $S$ is defined in equation~\eqref{eq:a1}. The constraint gives the condition that the connections are flat $F=dA+A\wedge A=0$. Imposing the constraint, using the remaining freedom to gauge transform the connection $A$ into the maximal torus of $\mathfrak{g}$, and imposing the canonical commutation relation, one obtains that the states are in one-to-one correspondence with points in the set~\cite{cs} described in equation~\eqref{eq:a2}.
%\begin{equation}
%\label{eq:b1}
%\frac{\Lambda_w}{W\ltimes q\Lambda_r}
%\end{equation}

In words, states are points on the weight lattice of $\mathfrak{g}$ with two states being identified if they differ by some combinations of the Weyl transformation and $q$ times the root lattice translation. We shall henceforth call this set the state set. Notice that because of the identification by $q\Lambda_r$, the number of states is finite, a fact that can also be seen from the compactness of the Chern-Simons phase space on $T^2$. In the large $q$ limit, the number of states is given by
\begin{equation}
\label{eq:b1}
\frac{q^r \det C}{\abs{W}}
\end{equation}
where $r$ is the rank of $\mathfrak{g}$, $C$ is the Cartan matrix, and $\abs{W}$ is the size of the Weyl group. This formula can either be derived from equation~\eqref{eq:a2} by noting that $\det C = \abs{\Lambda_w/\Lambda_r}$ or going back to canonical quantization and demanding that the phase space contains one state per $2\pi$ cell (we set $\hbar=1$).

\begin{figure}[htbp]
\centering
\includegraphics[scale=0.5]{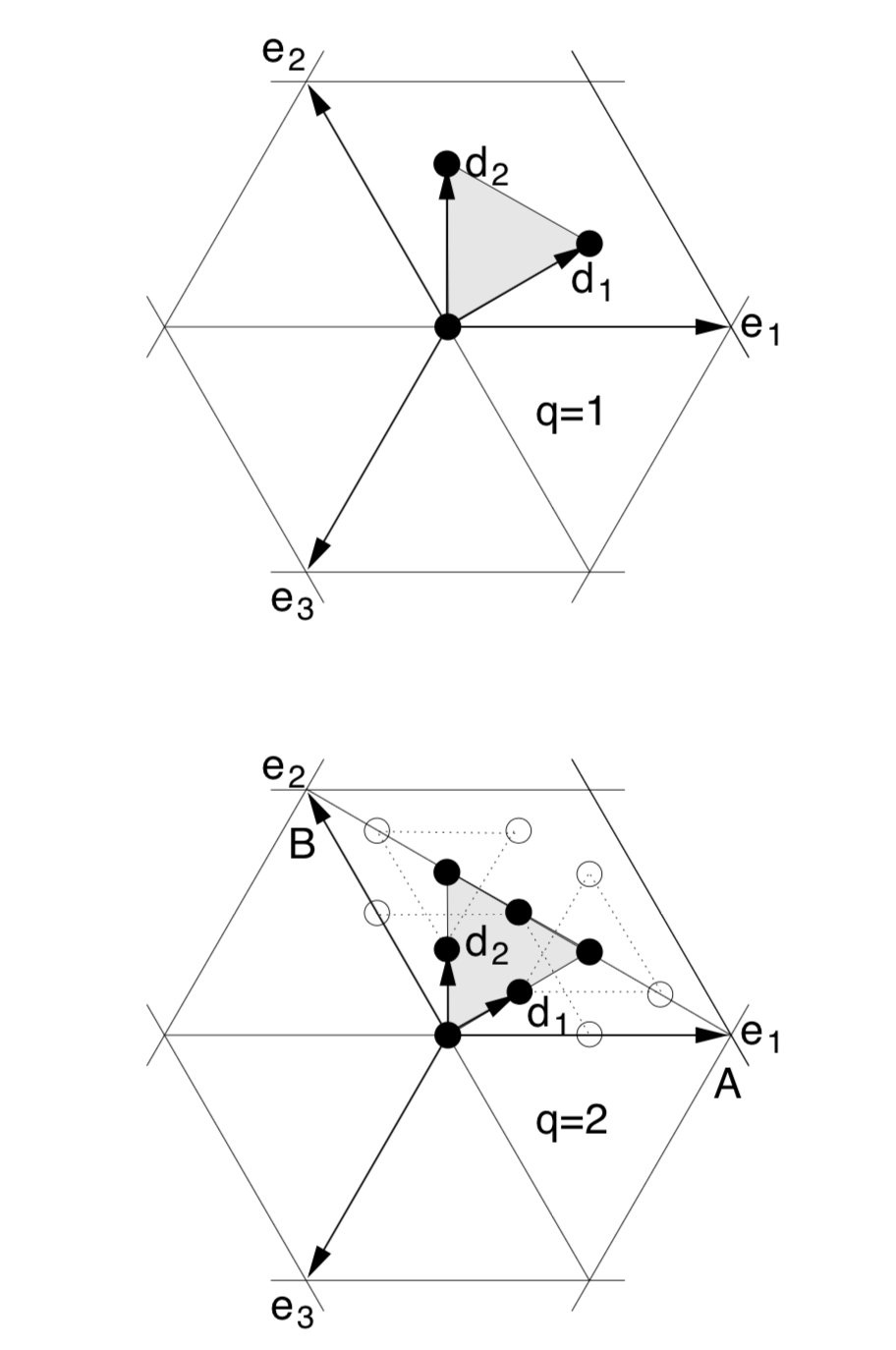}
\caption{In both pictures, $e_1$ and $e_2$ are the simple roots, and $d_1$ and $d_2$ are the fundamental weights. In the top picture, three black dots represent the three unique states of the level $q=1$ theory. In the bottom picture, the six black dots are the unique states of the level $q=2$ theory. It is easy to convince oneself that one can reach the white dots or other weight lattice points through a combination of Weyl reflections and $q$ times the root lattice translation on the black dots. Figure retrieved from~\cite{wenandzee}. \label{fig:1}}
\end{figure}

To give an example of the state set of $\mathfrak{su}(3)$ Chern-Simons theory at level $q=1$ and $q=2$, we present the figure (see Fig.~\ref{fig:1}) from ~\cite{wenandzee}\footnote{In~\cite{wenandzee}, a more pedestrian way of quantizing the Chern-Simons theory on $T^2$ for is used.}.

Fig.~\ref{fig:1} shows the unique states in the Hilbert space for the level 1 and the level 2 theory, respectively. Notice that the pattern continues to all levels: all states lie in the fundamental Weyl chamber, and that as the level increases, the number of state increases quadratically (roughly as the area of the shaded region). In fact, the number of states for the $\mathfrak{su}(3)$ theory at level $q$ is
\begin{equation}
\label{eq:b2}
\frac{q(q+1)}{2}
\end{equation}

This formula can be derived as follows. For simply laced algebra (which is our only focus in this paper), the root lattice $\Lambda_r$ is the same as the coroot lattice $\Lambda_{cr}$. A theory in affine Lie algebra shows that the state set~\eqref{eq:a2} is simply the highest weight representation of the corresponding affine Lie algebra $\tilde{\mathfrak{g}}$ at level $q$~\cite{kac}. Let a state be labeled by the Dynkin label $(a_1, a_2,...,a_r)$, where each $a_i$ is some nonnegative integer. The highest weight states satisfy the inequality
\begin{equation}
\label{eq:new0}
\sum_{i=1}^r c_i a_i \leq q
\end{equation}
where $c_i\in \mathbb{Z}_{>0}$ is the coefficient multiplying the $i$th simple root $\alpha_i$ in the expression for the highest root $\theta$:
\begin{equation}
\theta \equiv \sum_{i=1}^r c_i \alpha_i
\end{equation}

For the case of $\mathfrak{su}(3)$, the numbers are $c_1=c_2=1$, so the number of Chern-Simons states at level $q$ is the same as the number of solutions to the inequality
\[
a_1 + a_2 \leq q
\]
and solving for this gives exactly the quadratic formula~\eqref{eq:b2}.

Although the example is given for $\mathfrak{su}(3)$, the reader should keep in mind the generalization of the picture to other ADE gauge algebras. In the next section, we look at a special class of states on the state lattice. 

\section{A Special Class of States}
\label{sec:3}
\subsection{The Geometry and the Counting Problem}
The special states we want to focus on are those that belong to the set
\begin{equation}
Q_q= \frac{\Lambda_w}{W\ltimes q\Lambda_r} \cap \Lambda_r
\end{equation}
namely states that also roots. The physical motivation for considering such states is mentioned in \cite{ganorandju} and is explained in the section~\ref{sec:8}. Following the $\mathfrak{su}(3)$ example in Fig.~\ref{fig:1}, at level 1 there is one state that belongs to $Q_q$, the state at the origin. At level 2 there are two states, the additional one being at the position $d_1+d_2$. \par
The question we pose is: given $\mathfrak{g}$ and level $q$, how many states are in $Q_q$? There are two equivalent formulations of this counting problem. The first formulation uses the Cartan matrix and it naturally leads to the concept of Ehrhart polynomials. The second (dual) formulation uses the inverse of the Cartan matrix, and it leads to representation theory and connections to string theory. The rest of the section focuses on the first formulation of the problem.

Let the Cartan matrix of $\mathfrak{g}$ be $C$. In the basis of Dynkin labels, the rows of $C$ give the Dynkin coefficients of the simple roots. Since we are dealing with simply laced Lie algebras, $C^T=C$, so that the columns of $C$ also give the representation of the simple roots. If $y\in Q_q$, $y$ being the Dynkin label (as a column vector) of some state in the theory, then $y\in \Lambda_r$ by definition, so that it can be represented as a linear combination of the simple roots:
\begin{equation*}
y= Cx
\end{equation*}
where $x\in \mathbb{Z}_{\geq 0}^r$ is a column of nonnegative integers. The nonnegativity of $x$ is because states in $Q_q$ are all dominant weights. What constraints does $y$ have to satisfy? The first constraint on $y$ is that the entries are nonnegative (note that $x\geq 0 $ does not imply $y\geq 0$). The second constraint is the level $q$ constraint in equation~\eqref{eq:new0}. In terms of $x$, the constraints read
\begin{align}
\sum_{j=1}^r C_{ij}x_j &\geq 0 \\
\sum_{i,j=1}^r c_iC_{ij}x_j &\leq q
\end{align}\par
Geometrically, the above constraints define a rational polytope, a polygon whose vertices have rational coordinates. The solutions to the above constraints are simply integer points contained inside the rational polytope. By adding slack variables, any rational polytope can be represented as a system of linear equations~\cite{beckandrobins}
\begin{equation}
Ax'=b
\end{equation}
for some matrix $A$, vector $b$, and unknowns $x'$. 

For our problem, we need $r+1$ slack variables $k_1, k_2,...,k_{r+1} \in \mathbb{Z}_{\geq 0}$, so that the system of inequalities reduces to the system of equalities:
\begin{align}
\sum_{j=1}^r C_{ij}x_j - k_i &= 0 \\
\sum_{i,j=1}^r c_iC_{ij}x_j + k_{r+1} &= q
\end{align}\par
Stacking the vector $x$ and $k$ into $x'=(x,k)$, we have the system of linear equation $Ax'=b$ where $A$ is $r+1$ by $2r+1$ and $b$ is $r+1$ by 1:
\begin{align}
\label{eq:c1}
A&=
\begin{pmatrix}
C_{11} & C_{12}  & \hdots &C_{1r}&  -1& 0 & 0 & \hdots & 0\\
C_{21} & C_{22} & \hdots & C_{2r} & 0 & -1 & 0 & \hdots & 0 \\
\vdots & \vdots  & \vdots & \vdots & \vdots &\vdots & \vdots & \vdots & \vdots  \\
C_{r1} & C_{r2} & \hdots & C_{rr} & 0 & 0 & 0 & -1 & 0 \\
\sum_i c_iC_{i1} & \sum_i c_iC_{i2}  & \hdots  & \sum_i c_iC_{ir}   & 0  & \hdots  & 0 & \hdots&1
\end{pmatrix} \\
\label{eq:c2}
b&=
\begin{pmatrix}
0 \\
0 \\
\vdots \\
0 \\
q
\end{pmatrix}
\end{align}
The problem of counting states in $Q_q$ is then transformed into counting the solutions to this system of linear equations.

\subsection{Ehrhart Polynomials}
What we did in the previous section is under the guise of Ehrhart polynomials~\cite{ehrhart}. They are generating functions for counting lattice points contained in polytopes. Let us formulate the theory following the notation of~\cite{beckandrobins}. Let a rational polytope $\mathcal{P}$ be specified by a system of linear equations with slack variables added:
\begin{equation}
\mathcal{P}=\{x\in\mathbb{R}^d_{\geq 0}: Ax=b \}
\end{equation}
for some integer valued matrix $A$ and integer valued vector $b$. One considers the $t$th dilate of $\mathcal{P}$, defined as 
\begin{equation}
t\mathcal{P}=\{x\in\mathbb{R}^d_{\geq 0}: Ax=tb \}
\end{equation}
where $t\in \mathbb{Z}_{>0}$. Let $L_{\mathcal{P}}(t)$ denote the number of lattice points contained in the $t\mathcal{P}$:
\begin{equation}
L_{\mathcal{P}}(t)= \#\{x\in \mathbb{Z}^d_{\geq 0}: Ax=tb\}
\end{equation}

\par
The Ehrhart polynomial associated to the polytope $\mathcal{P}$ is defined as
\begin{equation}
\text{Ehr}_\mathcal{P}(z)  = 1 + \sum_{t\geq 1} L_{\mathcal{P}}(t) z^t
\end{equation}
\par
In our problem, the polytope is given as the $q$th dilate of 
\[
Ax' = 
\begin{pmatrix}
0 \\
0 \\
\vdots \\
0 \\
1
\end{pmatrix}
\]
with $A$ and $x'$ defined in the previous section (see equation~\eqref{eq:c1} and equation~\eqref{eq:c2}). We call this base polytope $\mathcal{Q}_\mathfrak{g}$, where the dependence on the Lie algebra $\mathfrak{g}$ is made explicit as each simply laced Lie algebra has its unique polytope. Since we are interested in computing the number of special states for each level $q>0$, we want to find the number of lattice points contained in the $q$th dilate of $\mathcal{Q}_\mathfrak{g}$ for each positive q. Therefore, the question posed in the previous section can be now phrased as finding the Ehrhart polynomial of $\mathcal{Q}_\mathfrak{g}$:
\[
\text{Ehr}_{\mathcal{Q}_\mathfrak{g}}(z)=?
\]

This concludes the first formulation of our problem. The dual formulation of the problem in terms of the inverse of Cartan matrices will be introduced in section~\ref{sec:5}. In the next section, we develop a formal method to compute $\text{Ehr}_{\mathcal{Q}_\mathfrak{g}}(z)$ in the most general way possible.

\section{Computation of $\text{Ehr}_{\mathcal{Q}_\mathfrak{g}}(z)$}
\label{sec:4}
In this section and the next, we compute $\text{Ehr}_{\mathcal{Q}_\mathfrak{g}}(z)$ for all simply laced $\mathfrak{g}$. We give two approaches to this computation. The first approach uses MacMahon's $\Omega$ operator method~\cite{macmahon} but quickly becomes tedious when the level $q$ becomes large. However, the merit of this approach is that it is general: it can be applied to solving for all Ehrhart polynomials given the constraints, and it will be amiss if we do not discuss the general solution. Due to the computational difficulty, we shall only use this approach to give the explicit formula for $\mathcal{Q}_\mathfrak{g}$ for the case $\mathfrak{g}=\mathfrak{su}(2)$. MacMahon's method is reviewed in section~\ref{sec:4.1} and the computation for $\mathfrak{su}(2)$ is done in section~\ref{sec:4.2}. 

The second formulation of the problem uses a hint from representation theory, and leads to the expression of $\text{Ehr}_{\mathcal{Q}_\mathfrak{g}}(z)$ in one full sweep. The second approach is inspired from a duality relation constructed from string theory (see section~\ref{sec:8} and ref~\cite{dijkgraaf}\cite{vafawitten}), without which it is not obvious how one can make a connection of $\text{Ehr}_{\mathcal{Q}_\mathfrak{g}}(z)$ to being solved by representation theory.

\subsection{The $\Omega$ Operator}
\label{sec:4.1}
In computing the number of ways of partitioning some integer $u$ into a sum of $n$ nonnegative integers $a_1+a_2+...+a_n=u$, the order of the integers in the sum does not matter. Therefore, the problem of counting n-partitions of $u$ is quite different from the problem of counting solutions to the the equation $a_1+a_2+...+a_n=u$, where the number of solutions is the coefficient of the $x^u$ term in the generating function
\begin{equation}
\frac{1}{(1-x)^n}
\end{equation}

To introduce the $\Omega$ operator, we focus on the number partition problem, and therefore we can assume an ordering of $a_i$ to be $a_1\geq a_2\geq...\geq a_n$ without loss of generality. One way to impose this ordering constraint is to consider the expression~\cite{macmahon}
\begin{equation}
\label{eq:d1}
\underset{\geq}{\Omega} \frac{1}{(1-\lambda_1 x) (1-\frac{\lambda_2}{\lambda_1} x)(1-\frac{\lambda_3}{\lambda_2}x)...(1-\frac{\lambda_{n}}{\lambda_{n-1}}x)}
\end{equation}
where the notation $\underset{\geq}{\Omega}$ means restricting terms that have only nonnegative powers of each $\lambda$ and setting each $\lambda$ to be one in the end. This is easily verified by expanding each fraction in power series.

The $\underset{\geq}{\Omega}$ leads to many identities. For example, one can again expand in power series and verify that
\begin{equation}
\underset{\geq}{\Omega}  \frac{1}{(1-\lambda x^{p_1})(1-\frac{x^{p_2}}{\lambda})} = \frac{1}{(1-x^{p_1})(1-x^{p_1+p_2})}
\end{equation}
Repeated use of this identity shows that equation~\eqref{eq:d1} is equal to 
\begin{equation}
\frac{1}{(1-x)(1-x^2)(1-x^3)...(1-x^n)}
\end{equation}
which is exactly the generating function to count the n-partition of some integer.

One can also compose the $\underset{\geq}{\Omega}$ operation. We modify the notation accordingly if there is any ambiguity with extra variables:
\begin{align*}
\underset{\lambda\geq}{\Omega}\underset{\mu\geq}{\Omega} \frac{1}{(1-\lambda\mu x) (1-\frac{y}{\lambda^2\mu})} &= 
\underset{\lambda\geq}{\Omega} \frac{1}{(1-\lambda x) (1-\frac{xy}{\lambda})} \\
&=\frac{1}{(1- x) (1-x^2y)} 
\end{align*}
This expression counts the number of partition into two nonnegative integers $a_1$ and $a_2$ such that the constraint $a_1 \geq 2a_2$ is satisfied\footnote{Reason: $\underset{\mu\geq}{\Omega}$ implements the condition that ordering does not matter, and $\underset{\lambda\geq}{\Omega}$ implements the constraint $a_1 \geq 2a_2$.}.

In fact, one can also have two other operators $\underset{\leq}{\Omega}$ and $\underset{=}{\Omega} $, defined in a self-explanatory way. As we shall see, we will be interested in identities involving $\underset{=}{\Omega}$. The three operators satisfy some useful algebraic relations listed in \cite{macmahon}, and can be used to compute $\underset{=}{\Omega}$. For example, let $F(\lambda)$ be some polynomial depending on $\lambda$ as the variable used in the $\Omega$ operators, the following expression
\begin{equation}
\underset{=}{\Omega} F(\lambda) = \underset{\geq}{\Omega} F(\lambda) + \underset{\geq}{\Omega}F(\lambda^{-1}) - F(1)
\end{equation}
can be used to derive identities involving $\underset{=}{\Omega}$ in terms of the identities involving $\underset{\geq}{\Omega}$. In the next subsection we will use an identity $\underset{=}{\Omega}$ to compute $\text{Ehr}_{\mathcal{Q}_\mathfrak{g}}(z)$ for $\mathfrak{g}=\mathfrak{su}(2)$. We will also sketch the idea of computing $\text{Ehr}_{\mathcal{Q}_\mathfrak{g}}(z)$ for a general simply laced $\mathfrak{g}$.

\subsection{Applying the $\Omega$ Operator}
\label{sec:4.2}
To illustrate the use of the $\Omega$ operator, we use it to write down a formal expression for the Ehrhart polynomial $\text{Ehr}_{\mathcal{Q}_\mathfrak{g}}(z)$. Recall that the base polytope is given by
\begin{equation}
Ax' = 
\begin{pmatrix}
0 \\
0 \\
\vdots \\
0 \\
1
\end{pmatrix}
\end{equation}
where $A$ ($r+1$ by $2r+1$) was given in equation~\eqref{eq:c1}. Let $\textbf{z}\equiv (z_1, z_2, ..., z_r, z)$. The Ehrhart polynomial is computed by the formal expression
\begin{equation}
\label{eq:d2}
\text{Ehr}_{\mathcal{Q}_\mathfrak{g}}(z)=\underset{z_1=}{\Omega}\underset{z_2=}{\Omega}...\underset{z_r=}{\Omega}\left(\prod_{i=1}^{2r+1}\frac{1}{1-z_1^{A_{1i}}z_2^{A_{2i}}...z_r^{A_{ri}} z^{A_{r+1,i}}}\right)
\end{equation}
where a composition of $r$ $\underset{=}{\Omega}$ is applied, each time restricting the polynomial to the $0$th order term of some $z_i$. This formula looks intimidating, but can be easily derived as follows. The term involving the product is simply the generating function for counting the combinations of $Ax'$. Imposing the constraint that the right hand side of the equation has $r$ vanishing entries means that one must restrict to the $0$th order term of $z_1,...z_r$. Since the last entry of the column on the right hand side is 1, the number of solutions to the $q$th dilate of the polytope is then the the coefficient of the $z^q$ term. This completes the argument that the formal expression in \eqref{eq:d2} computes the Ehrhart polynomial.

It is nice to have a formal expression like eqn.~\eqref{eq:d2} for the Ehrhart polynomial. If one wants to compute the first few terms, a computer can easily do the job. However, we want to take a step further and obtain a closed form solution, which as we shall see exists for all simply laced $\mathfrak{g}$.\par
To illustrate how one might obtain a closed form solution using the current formalism, we first focus on the case of $\mathfrak{su}(2)$. The Cartan matrix for $\mathfrak{su}(2)$ is simply 2,  the coefficient for the highest root is $c_1=1$. Therefore the $A$ matrix is 
\begin{equation}
A=\begin{pmatrix}
2 & -1 & 0 \\
2 & 0 & 1
\end{pmatrix}
\end{equation}

Using expression~\eqref{eq:d2}, the Ehrhart polynomial is given by
\begin{equation}
\text{Ehr}_{\mathcal{Q}_\mathfrak{su}(2)}(z)=\underset{z_1=}{\Omega}\left(\frac{1}{1-z_1^2 z^2} \frac{1}{1-z_1^{-1}} \frac{1}{1-z}\right)
\end{equation} 

Using the $\Omega$ operator identity \cite{macmahon}
\begin{equation}
\label{eq:d3}
=\underset{\lambda=}{\Omega} \frac{1}{(1-\lambda^2 x)(1-y\lambda^{-1} )}= \frac{1}{1-xy^2}
\end{equation}
we find that the Ehrhart polynomial for $\mathfrak{g}=\mathfrak{su}(2)$ is
\begin{equation}
\label{eq:d4}
\text{Ehr}_{\mathcal{Q}_\mathfrak{su}(2)}(z)= \frac{1}{(1-z)(1-z^2)}
\end{equation} 

Now let us generalize to arbitrary $\mathfrak{g}$. The first thing to note is that one needs more general $\Omega$ identities in addition to equation~\eqref{eq:d3}, since a variable $Z_i$ can appear more than two times in the product in equation~\eqref{eq:d2}. However, the $\Omega$ identities for a general polynomial are not documented and do not have clean solutions. This problem has been solved by~\cite{andrews}, where the authors developed the Omega Package in Mathematica to compute $\Omega$ identities for a general polynomial 
\begin{equation}
\frac{P(x_1,...,x_n;\lambda_1,...\lambda_r)}{\prod_{i=1}^n (1-x_i \lambda_1^{v_1(i)}...\lambda_r^{v_r(i)} )}
\end{equation}

For our problem, we can apply the program $r$ times to eliminate $z_1,...z_r$ to find $\text{Ehr}_{\mathcal{Q}_\mathfrak{g}}(z)$ of rank $r$. In the next section, we will use a trick from representation theory to compute all $\text{Ehr}_{\mathcal{Q}_\mathfrak{g}}(z)$ by hand, bypassing the need for the computer program computation.

\section{$\text{Ehr}_{\mathcal{Q}_\mathfrak{g}}(z)$ From Representation Theory}
\label{sec:5}
To connect $\text{Ehr}_{\mathcal{Q}_\mathfrak{g}}(z)$ to representation theory, we briefly summarize the McKay correspondence~\cite{mckay}, which associates any simply laced $\mathfrak{g}$ to some discrete subgroup $\Gamma(\mathfrak{g})$ of $SU(2)$. More specifically, 
\begin{align}
\Gamma(\mathfrak{su}(N)) &= \mathbb{Z}_{N} \\
\Gamma(\mathfrak{so}(2(N+2))) &= \text{Dic}_{N} \\
\Gamma(\mathfrak{e}_i) &= E_i \qquad\text{i=6,7,8}
\end{align}
where $\mathbb{Z}_N$ is the cyclic group of order $N$, $\text{Dic}_{N}$ is the dicyclic (binary dihedral) group of order $4N$, and $E_6, E_7, E_8$ are the binary tetrahedral group, binary octahedral group, and the binary icosahedral group, respectively (also called $2T$, $2O$, and $2I$).

For some simply laced $\mathfrak{g}$, consider the homomorphism $\Gamma(\mathfrak{g}) \to SU(q)$, or, in other words, $SU(q)$ representations of the group $\Gamma(\mathfrak{g})$. Let the number of Weyl-inequivalent representations\footnote{This means that two diagonal $SU(q)$ matrices are identified if the diagonal elements differ by some permutation.} $\Gamma(\mathfrak{g}) \to SU(q)$ be $b_q$, so that one forms the generating function to count the number of $SU(q)$ representations
\begin{equation}
\label{eq:e0}
\Phi_\mathfrak{g}(z) \equiv 1+\sum_{i=1} b_i z^i
\end{equation} 

We claim that this generating function is exactly the same as the Ehrhart polynomial corresponding to the same Lie algebra:

\begin{equation}
\label{eq:e1}
\text{Ehr}_{\mathcal{Q}_\mathfrak{g}}(z) = \Phi_\mathfrak{g}(z)
\end{equation}

This beautiful equation connects geometry (left hand side) with representation theory (right hand side) via the McKay correspondence and string theory (the equality sign). The rest of this subsection is devoted to proving this equality and using this equality to compute the closed form $\text{Ehr}_{\mathcal{Q}_\mathfrak{g}}(z)$ for $\mathfrak{g}=\mathfrak{su}(N)$ and $\mathfrak{g}=\mathfrak{so}(2(N+2))$. The formulae for the exceptional Lie algebras are computed in the appendix of \cite{ganorandju}. In section~\ref{sec:8}, we will ``prove'' this equation using a construction in string/M theory.

The strategy we use here is to look at the dual formulation by starting with weights, expressing them in terms of the simple roots, and imposing the constraints. This reverse process is carried out using the inverse the Cartan matrix. Since the inverse of Cartan matrices in general has fractional entries, this dual formulation is less suited for geometric arguments we had in the previous sections. Instead, we will use purely algebraic arguments to prove equation~\eqref{eq:e1}.

Let $\mathfrak{g}=\mathfrak{su}(r+1)$ and the corresponding Cartan matrix be $C$. Start with some weight $x$ in the state set~\eqref{eq:a2} which we reproduce here for convenience
\begin{equation}
\frac{\Lambda_w}{W\ltimes q\Lambda_r}
\end{equation}

We can represent $x$ as a vector in $\mathbb{Z}^r_{\geq 0}$ by using its Dynkin label $(x_1,...,x_r)$. It satisfies the constraint
\begin{equation}
\sum_{i=1}^r x_i \leq q
\end{equation}
as argued in section~\ref{sec:2}. \par

Since the simple roots span $\mathbb{R}^r$, the weight $x$ has a unique expansion in terms of the simple roots $\alpha_1, ... ,\alpha_r$:
\begin{equation}
x = \sum_{i=1}^r l_i \alpha_i
\end{equation} 
The expansion coefficients $l_j$ are simply given by multiplying $x$ (as a vector) by the inverse of the Cartan matrix\cite{kac}
\begin{equation}
l_i = \sum_{j=1}^r C^{-1}_{ij} x_j
\end{equation}\par
Since we want $x$ to lie on the root lattice $\Lambda_r$ as well, the only other constraint we need to impose is that each $l_j$ is an integer
\begin{equation}
\label{eq:new1}
l_i = \sum_{j=1}^r C^{-1}_{ij} x_j \in \mathbb{Z}
\end{equation}
\par
The inverse of the Cartan matrix for $\mathfrak{su}(N)$ algebra has an interesting modular structure. Here, we display $C^{-1}$ for $\mathfrak{su}(4)$ and $\mathfrak{su}(7)$:
\begin{align}
C^{-1}_{\mathfrak{su}(4)} &=
\frac{1}{4}
\begin{pmatrix}
3 & 2 & 1 \\
2  & 4 & 2 \\
1 & 2 & 3
\end{pmatrix} \\
C^{-1}_{\mathfrak{su}(7)} &=
\frac{1}{7}
\begin{pmatrix}
6 & 5 & 4 & 3 & 2 & 1\\
5 & 10 & 8 & 6 & 4 & 2 \\
4 & 8 & 12 & 9 & 6 & 3 \\
3 & 6 & 9 & 12 & 8 & 4 \\
2 & 4 & 6 & 8 & 10 & 5 \\
1 & 2 & 3 & 4 & 5 & 6 \\
\end{pmatrix}
\end{align}\par
In general, the formula for $C^{-1}_{\mathfrak{su}(N)}$ is~\cite{dynkin}
\begin{equation}
C^{-1}_{\mathfrak{su}(N), ij} = \frac{1}{N} \left[\min(i,j)\times N - ij\right]
\end{equation} 

Even though we have $r$ number of constraints from equation~\eqref{eq:new1}, we shall see that because of the peculiar property of $C^{-1}_{\mathfrak{su}(N)}$, there is effectively only one constraint, the one imposed by the last row of the matrix:
\begin{equation}
\label{eq:e2}
\frac{1}{N} \sum_{i=1}^{N-1} ix_i \in \mathbb{Z}
\end{equation} 

A quick proof that constraint~\eqref{eq:e2} implies the rest of the constraints is as follows. The matrix elements $C^{-1}_{\mathfrak{su}(N),ki}$ of the $k$th row are
\begin{align*}
C^{-1}_{\mathfrak{su}(N),ki} =\begin{cases} 
\frac{(N-k)i}{N} \qquad k> i \\ 
\frac{(N-i)k}{N} \qquad k\leq i 
\end{cases} 
\end{align*}\par
Using this, we obtain the constraint imposed by the $k$th row:
\begin{align*}
l_k &= \sum_{j=1}^{N-1} C^{-1}_{\mathfrak{su}(N), kj} x_j \\
&= \sum_{i=1}^{k-1}  \frac{(N-k)i}{N} x_i +  \sum_{i=k}^{N-1} \frac{(N-i)k}{N} x_i \\
&= -\frac{k}{N} \sum_{i=1}^{N-1} ix_i \mod 1
\end{align*} \par
Up to integers, the constraint imposed by the $k$th row is simply $k$ times that of the constraint imposed by the last row. Therefore, the only unique constraint in \eqref{eq:new1} is the one given by the last row. In summary, there are two constraints on $x$:
\begin{align}
\label{eq:e3}
\sum_{i=1}^{N-1} x_i \leq q \\
\label{eq:e4}
\frac{1}{N} \sum_{i=1}^{N-1} ix_i \in \mathbb{Z}
\end{align}
We would now like to argue that this is exactly the same constraints satisfied by $SU(q)$ representations of $\Gamma(\mathfrak{su}(N))=\mathbb{Z}_N$.

\subsection{$SU(q)$ representations of $\Gamma(\mathfrak{su}(N))=\mathbb{Z}_N$}
Since finite dimensional representations are built up from irreducible representations, we first look at the irreducible representations of $\mathbb{Z}_N$. There are in total $N$ 1-dimensional irreducible representations. The $k$th irreducible representation is given by the $k$th power of the $N$th root of unity 
\begin{equation}
\omega_k = \exp\left( \frac{2\pi i k}{N}\right)
\end{equation}\par
Here, $k$ runs from 0 to $N-1$, with $k=0$ being the trivial representation. The $SU(q)$ representations are constructed by inserting $x_k$ copies of the $k$th irreducible representation. By definition, the determinant of any $SU(q)$ matrix must be 1, so we have 
\begin{equation}
\prod_{k=0}^{N-1} \omega_k^{x_k} = \exp\left(\frac{2\pi i }{N} \sum_{k=0}^{N-1} k x_k\right) =1
\end{equation}
But this is equivalent to the constraint~\ref{eq:e4}:
\begin{equation*}
\frac{1}{N} \sum_{i=1}^{N-1} ix_i \in \mathbb{Z}
\end{equation*}

The constraint that the dimensions of the irreducible representations add up to $q$ is
\begin{equation}
\sum_{i=0}^{N-1} x_i = q
\end{equation}
However, treating $x_0$ as a slack variable, this constraint is equivalent to~\eqref{eq:e3}
\begin{equation*}
\sum_{i=1}^{N-1} x_i \leq q
\end{equation*}

Therefore, we have shown that the two counting problems satisfy the same constraints and are secretly one and the same. This concludes the proof of eqn.~\eqref{eq:e1} for the case $\mathfrak{g}=\mathfrak{su}(N)$:
\begin{equation*}
\text{Ehr}_{\mathcal{Q}_\mathfrak{g}}(z) = \Phi_\mathfrak{g}(z)
\end{equation*}
 
In the next subsection, we give explicit formulae for $\text{Ehr}_{\mathcal{Q}_\mathfrak{g}}(z) $ by computing $\Phi_\mathfrak{g}(z)$.
 
\subsection{Computation of $\Phi_\mathfrak{g}(z)$}
We want to find the generating function~\eqref{eq:e0}
\begin{equation*}
\Phi_\mathfrak{g}(z) = 1+ \sum_i b_i z^i
\end{equation*}
in which $b_q$ counts the number of inequivalent $SU(q)$ representations of $\mathbb{Z}_N$. We saw in the previous subsection that the $N$ irreducible representations are given by the $N$th roots of unity. Consider the function\footnote{We thank O. Ganor for pointing out this trick.}
\begin{equation}
\frac{1}{(1-z)(1-wz)(1-w^2z)...(1-w^{N-1}z)}
\end{equation} \par
The coefficient of $z^q$ is in general a sum of $n$ terms ($n$ is some integer), representing $n$ ways of building a $q$-dimensional representation. Each of the $n$ terms has some coefficient $w^m$ for some integer $m\in\mathbb{Z}$, which represents the determinant of that representation. We want to retain terms of determinant 1 only. To project out terms of non-unit determinant, we simply have to sum over $w$ in $w^m$ and divide by $N$, since we know from Fourier analysis that 
\[
\sum_w w^{m} =\begin{cases}
N \qquad m=0 \mod N \\
0 \qquad m\neq 0 \mod N
\end{cases}
\]
where the sum is taken over the $N$th root of unity. Therefore, the generating function is
\begin{equation}
\label{eq:zkpoly}
\Phi_{\mathfrak{su}(N)}(z)  =1+ \sum_i b_i z^i= \frac{1}{N}\sum_{i=0}^{N-1} \frac{1}{(1-z)(1-w^iz)(1-w^{2i}z)...(1-w^{(N-1)i}z)}
\end{equation}
\par
One can check that $b_i$ is a quasi-polynomial in $i$, a property shared by Ehrhart polynomial. Finally, we use the theorem proved in the last subsection to give the explicit expression for the Ehrhart polynomial
\begin{equation}
\label{eq:zkpoly2}
\text{Ehr}_{\mathcal{Q}_\mathfrak{su}(N)}(z) = \frac{1}{N}\sum_{i=0}^{N-1} \frac{1}{(1-z)(1-w^iz)(1-w^{2i}z)...(1-w^{(N-1)i}z)}
\end{equation}

For example, let $\mathfrak{g}=\mathfrak{su}(2)$. Then 
\begin{align*}
\text{Ehr}_{\mathcal{Q}_\mathfrak{su}(2)}(z)  &= \frac{1}{2} \left(\frac{1}{(1-z)(1-z)} + \frac{1}{(1-z)(1+z)}\right) \\
&=\frac{1}{(1-z)(1-z^2)}
\end{align*}
which agrees with what we obtained in equation~\eqref{eq:d4} using the $\Omega$ operator calculus.

\section{Ehrhart Polynomial for $\mathfrak{so}(2(N+2))$}
\label{sec:6}
In the previous subsections, we obtained the Ehrhart polynomial for $\mathfrak{su}(N)$ by counting the $SU(q)$ representations of $\mathbb{Z}_N$. We now use the same method to compute the Ehrhart polynomial for $\mathfrak{so}(2(N+2))$, $N\geq 1$. Here, $N$ is shifted by 2 because of convenience. The Dynkin diagram associated to $\mathfrak{so}(2(N+2))$ is $D_{N+2}$. According to the McKay correspondence, we should be looking for the representation of the discrete group $\text{Dic}_{N}$, the dicyclic (or binary dihedral) group of order $N$. Since $\text{Dic}_{N}$ is less well-known than $\mathbb{Z}_N$, we analyze the group structure in detail and derive the irreducible representations in the next subsection. After that, we will use the irreducible representations to construct the $SU(q)$ representation of $\text{Dic}_{N}$ and obtain the generating function $\Phi(z)$. We prove that, in a spirit similar to what we did for the $\mathfrak{su}(N)$ case, the generating function $\Phi(z)$ coincides with the Ehrhart polynomial for $\mathfrak{so}(2(N+2))$.

\subsection{$SU(q)$ representation of $\text{Dic}_N$}
The group $\text{Dic}_{N}$ is defined by the following multiplication rules:
\begin{align*}
r^{2N} &= e \\
s^2 &= r^N \\
s^{-1} r s &= r^{-1}
\end{align*}
where $e$ is the identity element. The reader may have noticed a similarity to the dihedral group of order $2N$, identifying $r$ with the fundamental rotation and $s$ with the rotation. The difference here is that the reflection $s$ does not square to the identity. Instead, it squares to a central element of the group. \par
\begin{figure}[htbp]
\centering
\begin{tikzpicture}
\begin{scope}[start chain]
\dnode{(1,1)}
\dnode{(2,2)}
\dnode{(3,2)}
\dydots
\dnode{(N,2)}
\dnoder{(N+1,1)}
\end{scope}
\begin{scope}[start chain=br going above]
\chainin(chain-2);
\dnodebr{(0,1)};
\end{scope}
\begin{scope}[start chain=br2 going above]
\chainin(chain-5);
\dnodebr{(N+2,1)};
\end{scope}
\end{tikzpicture}
\caption{The affine Dynkin diagram for $D_{N+2}$. The tuple $(x,y)$ indicates the $x$th simple root with (co)mark $y$. The mark is the same as the comark here because the algebra is simply-laced. By the McKay correspondence, the mark also indicates the dimension of the corresponding irreducible representation. Therefore, there are four 1-dimensional irreducible representations and $N-1$ 2-dimensional irreducible representations. \label{fig:2}}
\end{figure}
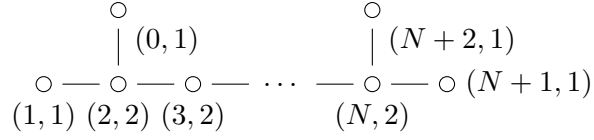

To analyze the irreducible representations of this group, we need to understand the conjugacy classes. For $\text{Dic}_{N}$, there are $N+3$ conjugacy classes:
\begin{equation}
\{e\},\{r,r^{-1}\},\{r^2,r^{-2}\},..., \{r^{N-1}, r^{-N+1}\},\{r^N\}, \{sr^{2k}\},\{sr^{2k+1}\}
\end{equation}
where for the last two conjugacy classes, $k$ is an integer running from 0 to $2N-1$. Therefore, there are $N+3$ irreducible representations. The affine Dynkin diagram of $D_{N+2}$ is shown in Fig.~\ref{fig:2}~\cite{kac}.

By the McKay correspondence~\cite{mckay} or by abelianizing the group, one sees that there are four 1-dimensional irreducible representations and $N-1$ 2-dimensional irreducible representations. The 1-dimensional irreducible representations for $\text{Dic}_{N}$ behave differently for $N$ even and $N$ odd. In the following, we restrict $N$ to be an even integer, since the case for $N$ odd can be treated analogously. We present the character table:
\begin{table}[htbp]
\centering
 
\begin{tabular}{|c|c|c|c|c|c|c|c|c|}
	\hline
	$\{e\}$ & $\{r,r^{-1}\}$ & $\{r^2,r^{-2}\}$ & ... &$\{r^{N-1}, r^{-N+1}\}$ & $\{r^N\}$ & $\{sr^{2k}\}$  & $\{sr^{2k+1}\}$ \\
	\hline 
	1 & 1& 1& ... & 1 & 1 & 1 & 1 \\
	1 & 1& 1& ... & 1 & 1 & -1 & -1 \\
	1 & -1& 1& ... & -1 & 1 &  1 & -1 \\
	1 & -1& 1& ... & -1 & 1 & -1 & 1 \\
	2 & $w+w^{-1}$ & $w^2+ w^{-2}$ & ... & $w^{N-1}+w^{-(N-1)}$ & -2 & 0 & 0 \\
	2 & $w^2+w^{-2}$ & $w^4+ w^{-4}$ & ... & $w^{2(N-1)}+w^{-2(N-1)}$ & 2 & 0 & 0 \\
	$\vdots$ & $\vdots$ & $\vdots$ & $\vdots$ & $\vdots$ & $\vdots$ & $\vdots$ & $\vdots$   \\
	2 & $w^{N-1}+w^{-N+1}$ & $w^{2(N-1)}+ w^{-2(N-1)}$ & ... & $w^{(N-1)^2}+w^{-(N-1)^2}$ & -2 & 0 & 0 \\
	\hline
	\end{tabular}
	\caption{Character table for the $\text{Dic}_N$ group. The first four lines are the characters for the four 1-dimensional irreducible representations. Note that the first line is the trivial 1-dimensional representation. The rest of the irreducible representations are 2-dimensional. $w$ represents the $2N$th root of unity $\exp(\pi i/N)$. One can check that the character orthogonality relation holds.}
\end{table}

For the $N-1$ 2-dimensional irreducible representations, we note that $r$ and $s$ take on the following form:t
\begin{align*}
r&=\begin{pmatrix}
\exp(m\pi i/N) & 0 \\
0 & \exp(-m\pi i/N) 
\end{pmatrix} \quad m=1,...,N-1 \\
s&=
\begin{cases}
\begin{pmatrix}
0 & 1 \\
-1 & 0
\end{pmatrix}, m \text{ odd} \\
\begin{pmatrix}
0 & 1 \\
+ 1 & 0
\end{pmatrix}, m \text{ even}
\end{cases} 
\end{align*}

In particular, for the 2-dimensional irreducible representations, $\det r =1$, and a non-unit determinant can only come from $s$. The formula for $s$ shows that the determinant of the 2-dimensional representations\footnote{In our case, we define the determinant of a representation as follows. If all conjugacy classes in the representation have determinant 1, we say that the representation has determinant 1. If some conjugacy classes have non-unit determinant, we pick the determinant $D$ that, in the polar decomposition, has the smallest angle $\theta\in [0, 2\pi)$ and call $D$ the determinant of the representation. For example, if $\omega$ is the 3rd root of unity, $\omega = \exp(2\pi i/3)$. Suppose there are two conjugacy classes with non-unit determinant, one with determinant $\omega$ and the other $\omega^2$. Because $\omega$ has a smaller polar angle than $\omega^2$, we say that the representation has determinant $\omega$. } alternates between 1 and -1. For example, $\text{Dic}_2$ has one 2-dimensional irreducible representation with determinant 1, whereas $\text{Dic}_4$ has three 2-dimensional irreducible representations, with determinant 1, -1, 1, respectively. \par
We now use the above information to compute the generating function for the $SU(q)$ representation of $\text{Dic}_N$. The $SU(q)$ representation is constructed out of the $N+3$ irreducible representations so that the dimensions add up to $q$. Let the number of the four 1-dimensional representations be $x_0,x_1,x_{N+1},x_{N+2}$, respectively\footnote{The numbering here is to make connection with the numbering of the nodes in the affine Dynkin diagram.}. Let the number of 2-dimensional representations be labeled by $x_2, x_3,..., x_N$. The constraint on the size of the representation is
\begin{equation}
\label{eq:f1}
x_0 + x_1 + 2x_2 + 2x_3 + ... + 2x_N + x_{N+1}+x_{N+2} = q
\end{equation}

In addition to the size constraint, we also have the unit-determinant constraint. From the character table, we see that a $-1$ determinant can only come from three columns\footnote{Note that the entries in the character table contain the trace of the conjugacy classes. The determinant coincides with the trace for 1-dimensional representations. For 2-dimensional representations the determinant in this case was computed earlier by looking at the explicit matrix representations. There are other columns that have $-1$ determinant, but one can show that there are only these three independent columns to consider.}: $\{r,r^{-1}\}$, $\{sr^{2k}\}$, and $\{sr^{2k+1}\}$. The unit-determinant constraints coming from the three columns are
\begin{align}
\label{eq:f2}
\frac{x_{N+1}+ x_{N+2}}{2} &\in \mathbb{Z} \\
\label{eq:f3}
\frac{x_1+x_3+x_5+...+x_{N-1}+x_{N+2}}{2} &\in \mathbb{Z} \\
\label{eq:f4}
\frac{x_1+x_3+x_5+...+x_{N-1}+x_{N+1}}{2} &\in \mathbb{Z} 
\end{align} 

Note that the constraints are not independent, since \eqref{eq:f3} and \eqref{eq:f4} imply \eqref{eq:f2}. Therefore, we effectively only have two determinant constraints. 

To write down the generating function $\Phi_{\mathfrak{so}(2(N+2))}$, we can use the independent constraints \eqref{eq:f3} and \eqref{eq:f4}. The idea is similar to the $\mathfrak{su}(N)$ case. Consider the function
\begin{equation}
\frac{1}{(1-z)(1-w_1 z)(1-w_2z)(1-w_1w_2 z)(1-z^2)^{N/2}(1-w_1w_2 z^2)^{N/2-1}}
\end{equation}

Here, $w_1, w_2 \in \{1,-1\}$. The first four terms represent the contributions of inserting the four 1-dimensional representations, and the last two terms represent the contributions of inserting the $N-1$ copies of the 2-dimensional representations. If we expand the fractions into a power series, a generic term would look like 
\begin{equation}
c w_1^{k_1}w_2^{k_2}z^n
\end{equation}
where $c, k_1, k_2 \in \mathbb{Z}$. The $w_2^{k_2}$ term represents the determinant contribution from the two 1-dimensional representations labeled by $N+1$ and $N+2$. We want to retain the term that satisfies $w_2^{k_2}=1$. This can be done by a projection similar to what we did for the $\mathfrak{su}(N)$ case, except now we need to sum over $\omega_2 \in \{1,-1\}$ and divide by 2. We also need to do a similar projection on $w_1$. In fact, the two projections help us retain the terms that satisfy the constraints \eqref{eq:f3} and \eqref{eq:f4}. Therefore, we need to compute
\[
\Phi_{\mathfrak{so}(2(N+2))}(z) =\frac{1}{4}\sum_{w_1,w_2} \frac{1}{(1-z)(1-w_1 z)(1-w_2z)(1-w_1w_2 z)(1-z^2)^{N/2}(1-w_1w_2 z^2)^{N/2-1}} 
\]e
which gives us the answer for $\Phi_{\mathfrak{so}(2(N+2))}(z)$:
\begin{equation}
\label{eq:dicrep}
\frac{1}{4}\left(\frac{1}{(1-z)^4(1-z^2)^{N-1}} + \frac{2}{(1-z^2)^2(1-z^2)^{N/2}(1+z^2)^{N/2-1}} +\frac{1}{(1-z^2)^2(1-z^2)^{N-1}} \right)
\end{equation}

As an example, we set $N=2$, so that we are looking at the generating function for $SU(q)$ representations of Dic$_2$. The first few terms of~\eqref{eq:dicrep} are
\[
\Phi_{\mathfrak{so}(8)}(z) = 1 + z + 5 z^2 + ...
\]
which suggests that there are one $SU(1)$ representation and five $SU(2)$ representations. The one $SU(1)$ representation is just the trivial representation. Even though there are four unitary 1-dimensional representations for Dic$_2$, only the trivial representation has unit determinant as can be seen from the character table. The reader can easily work out the five $SU(2)$ representations, one of which comes from the unique \textit{irreducible} 2-dimensional representation of Dic$_2$. The rest come from the \textit{reducible} 2-dimensional representations by combining the 1-dimensional irreducible representations such that the determinant constraint is satisfied.

We will show in the next subsection that this is exactly the Ehrhart polynomial $\text{Ehr}_{\mathcal{Q}_\mathfrak{so}(2(N+2))}(z)$  for the ${D}_{N+2}$ polytope defined in a similar fashion as in section~\ref{sec:3}.

\subsection{Equivalence to the Ehrhart polynomial}
We are looking at the level $q$ highest weight representations of $\mathfrak{so}(2(N+2))$ which also lie on the root lattice. To show that the Ehrhart polynomial coincides with the generating function obtained in the previous subsection, we simply show that the two systems have the same constraints. Let the Dynkin label of some highest weight representation be $\lambda= (x_1,x_2,...,x_{N+2})$, each term some nonnegative integer. The level $q$ constraint yields
\begin{equation}
\label{eq:f5}
(\lambda,\theta) = x_1+ 2x_2 + 2x_3 + ... + 2x_{N} + x_{N+1} +x_{N+2} \leq q
\end{equation}
where $\theta$ is the highest root whose expansion coefficients in terms of the simple roots can be read off from the affine Dynkin diagram. By adding a slack variable to turn the inequality into an equality, we reproduce the first constraint \eqref{eq:f2}. Next, we need to demand that the weights are expressed as integer combinations of simple roots:
\begin{equation}
\label{eq:new2}
\sum_{j=1}^{N+2} C^{-1}_{ij}x_j \in \mathbb{Z}
\end{equation}
The inverse of the Cartan matrix for the D-series has an interesting form~\cite{weiandzou}. Since the matrix is symmetric, we only give the values for the upper half of the matrix. Let $C^{-1}$ be the inverse of the Cartan matrix for $D_{N+2}$. We have
\[
C^{-1}_{ij} = \begin{cases}
i, & 1\leq i \leq j \leq N \\
i/2, & i\leq N, j=N+1 \ \text{or} \ N \\
\frac{N}{4}, & i=N+1, j=N+2 \\
\frac{N+2}{4}, & i=j=N+1 \ \text{or} \ N+2
\end{cases}
\] 
\par
We give an example of $C^{-1}_{ij}$ for $D_6$ where $N=4$:
\[
C^{-1}_{\mathfrak{so}(12)} = 
\begin{pmatrix}
1 & 1 & 1 & 1 & 1/2 & 1/2 \\
1 & 2 & 2 & 2 & 1 & 1\\
1 & 2 & 3 & 3 & 3/2 & 3/2 \\
1 & 2 & 3 & 4 & 2 & 2 \\
1/2 & 1 & 3/2 & 2 & 3/2 & 1 \\
1/2 & 1 & 3/2 & 2 & 1 & 3/2 \\
\end{pmatrix}
\]\par
Note that the $N$ by $N$ block is integer-valued, so it does not enter into the constraint \eqref{eq:new2}. We only have to worry about the last two columns and the last two rows (which are the same as the last two columns by symmetry). For our case, $N$ is an even number, so the only fraction that enter into the constraints modulo 1 is $1/2$. Restricting our attention to the last two columns, we see that as the row number increases from $1$ to $N$, the values of the last two columns alternate between being half-integer valued and integer valued. Therefore, from the first $N$ rows of $C^{-1}$, we effectively get only one constraint:
\begin{equation}
\label{eq:f6}
\frac{x_{N+1}+x_{N+2}}{2} \in \mathbb{Z}
\end{equation}\par

The constraints coming from the last two rows can be deduced similarly. Restricting ourselves to the last two rows, as the column number $j$  increases from 1 to $N$, the values of $C^{-1}_{N+1,j}$ and $C^{-1}_{N+2,j}$ alternate between being half-integer valued and integer valued. Taking into account of the last two columns, the constraints are
\begin{align}
\label{eq:f7}
\frac{x_1+x_3+x_5+...+x_{N-1}+x_{N+2}}{2} &\in \mathbb{Z} \\
\label{eq:f8}
\frac{x_1+x_3+x_5+...+x_{N-1}+x_{N+1}}{2} &\in \mathbb{Z} 
\end{align}\par
We see that \eqref{eq:f5}, \eqref{eq:f7}, and \eqref{eq:f7} are exactly the same constraints we obtained in the last subsection. This establishes the equivalence of the generating function for $SU(q)$ representation of $\text{Dic}_{N}$ group and the Ehrhart polynomial for the $\mathfrak{so}(2(N+2))$ polytope. In fact, one can repeat the same argument to show that the equivalence holds for the exceptional Lie algebras as well. This concludes the proof of remarkable formula \eqref{eq:e1}.

\section{A New Perspective on the McKay Correspondence}
\label{sec:7}
Let $C$ be some Cartan matrix of an ADE type Lie algebra of rank $r$. Let $G_C$ be the discrete $SU(2)$ subgroup corresponding to $C$ according to the McKay correspondence. Define $[C^{-1}]$ as $C^{-1}$ modulo 1, where modulo 1 is done element-wise. Our computation in the previous two sections shows that a great deal of information is hidden in this object. In particular, $[C^{-1}]$ can tell us about the determinant\footnote{The determinant of a representation is defined in footnote 7.} of the irreducible representations of the group $G_C$. To make it more precise, we define the following $\vee$ operator acting on the rational numbers in the \textit{congruence class} of 1 as
\[
a \vee b =
\begin{cases}
a + b, & \text{if} \ a=0 \ \text{or} \ b=0 \\
a, &  \text{$b$ is a nonzero integer multiple of $a$} \\
b, & \text{$a$ is a nonzero integer multiple of $b$}
\end{cases}
\]\par
Note that this definition comes with a priority structure: there could be cases where condition 2 and 3 are both satisfied. In that case, we stick with condition 2 and demand that $a\vee b= a$. For example, bearing in mind that we are working within the congruence class of 1, the above rules imply
\begin{align*}
2/3 \vee 1/3 &= 2/3 \quad\text{because $1/3$ is 2 times $2/3$}\\
5/7 \vee 6/7 &= 5/7 \quad\text{because $6/7$ is 4 times $5/7$ }\\
1/4 \vee 0 &= 1/4 \\
1/2 \vee 1/2 &=1/2
\end{align*} \par

As we shall see, we never have to worry about the case when $a$ and $b$ do not satisfy the three cases. To use the $\vee$ operator, let us define $X_i$ to be the $i$th row of $[C^{-1}]$. Let the operator $\vee$ act element-wise on $X_i$. Define the row vector $X$ as
\begin{equation}
\label{eq:g1}
X \equiv X_1 \vee X_2 \vee ... \vee X_r
\end{equation}
The dual group $G_C$ has $r+1$ irreducible representations, among which 1 of them is the trivial representation. We claim that the determinant of the rest of the $r$ non-trivial representations, encoded in the row vector $D=(d_1,...d_r)$, can be found by 
\begin{equation}
\label{eq:g2}
D = \exp 2\pi i X
\end{equation}
where the exponential is taken element-wise on $X$, producing another row vector. 

This claim can be proved by checking for all ADE Lie algebras. The information presented in the previous sections is enough for the readers to check the claim for the $A$ and the $D$ series. Here we will focus on the exceptional series $\mathfrak{e}_6$, $\mathfrak{e}_7$, and $\mathfrak{e}_8$. The character tables for the dual group are worked out in~\cite{ganorandju}. We give a brief summary. 
\begin{itemize}
\item $\mathfrak{e}_6$. The dual group has six nontrivial irreducible representations, of which two have determinant $\exp(2\pi i/3)$ and two have determinant $\exp(4\pi i/3)$. The other two have determinant 1. 
\item $\mathfrak{e}_7$. The dual group has seven nontrivial irreducible representations, of which three have determinant $-1$ while the rest has determinant 1. 
\item $\mathfrak{e}_8$. All representations of the dual group have unit determinant.
\end{itemize}

Let us now compare the prediction made by equation~\eqref{eq:g2} to the facts cited above.
 
The most trivial case to check is $\mathfrak{e}_8$, whose inverse Cartan matrix modulo 1 vanishes. In this case, $X_1=X_2=...=X_8=0$, so according to the formula above, all irreducible representations of the binary icosahedral group have unit determinant. \par
For $\mathfrak{e}_7$, $[C^{-1}]$ is
\[
[C^{-1}_{\mathfrak{e}_7}]=
\begin{pmatrix}
0 & 0 & 0 & 0 & 0 & 0 & 0 \\
0 & 0 & 0 & 0 & 0 & 0 & 0 \\
0 & 0 & 0 & 0 & 0 & 0 & 0 \\
0 & 0 & 0 & 1/2 & 0 & 1/2 & 1/2 \\
0 & 0 & 0 & 0 & 0 & 0 & 0 \\
0 & 0 & 0 & 1/2 & 0 & 1/2 & 1/2 \\
0 & 0 & 0 & 1/2 & 0 & 1/2 & 1/2 
\end{pmatrix}
\]\par
The $\vee$ sum of all the rows gives $(0,0,0,1/2,0,1/2,1/2)$. Equation~\eqref{eq:g2} shows that the determinants are $(1,1,1,-1,1,-1,-1)$, agreeing with the facts above.\par
For $\mathfrak{e}_6$, $[C^{-1}]$ is
\[
[C^{-1}_{\mathfrak{e}_6}]=
\begin{pmatrix}
1/3 & 2/3 & 0 & 1/3 & 2/3 & 0 \\
2/3 & 1/3 & 0 & 2/3 & 1/3 & 0  \\
0 & 0 & 0 & 0 & 0 & 0  \\
1/3 & 2/3 & 0 & 1/3 & 2/3 & 0 \\
2/3 & 1/3 & 0 & 2/3 & 1/3 & 0 \\
0 & 0 & 0 & 0 & 0 & 0\\
\end{pmatrix}
\]\par
The $\vee$ sum of all the rows gives $(1/3,2/3,0,1/3,2/3,0)$. Equation~\eqref{eq:g2} shows that the determinants are $(w, w^2, 1, w, w^2, 1)$, where $w$ is the third root of unity. This also agrees with the facts cited above, and completes the proof of equation~\eqref{eq:g2}. \par

\section{The Physical Origin of the Duality}
\label{sec:8}
In this section, we give an overview of the physics underlying equation~\eqref{eq:e1}. As we shall see in this section, the right hand side of equation~\eqref{eq:e1} will be interpreted as the generating function for computing the dimension of the ground state Hilbert space of $\mathcal{N}=4$ $SU(q)$ Yang-Mills theory on $S^3/\Gamma$. The left hand side of equation~\eqref{eq:e1} is the generating function for computing the dimension of a certain subspace~\eqref{eq:a3} of the Hilbert space of level $q$ Chern-Simons theory with gauge algebra $\mathfrak{g}(\Gamma)$, where $\mathfrak{g}(\Gamma)$ is given by the McKay correspondence. The SYM theory ground states are simply the flat $SU(q)$ Wilson lines wrapping around the $\Gamma$ singularity, which in the language of representation theory translates to $SU(q)$ representation of $\Gamma$ up to conjugation and Weyl group identification. To prove~\eqref{eq:e1}, we would like to show that these two Hilbert spaces indeed have the same dimension for a given $SU(q)$ gauge group on the SYM side and for any ADE singularity $\Gamma$.

To show this, consider in IIB string theory a stack of $N$ D3 branes that in the Euclidean signature span the 0123 directions. The ADE singularity $\Gamma$ acts on the world volume\footnote{Note that this setup is different from the D-brane on orbifold setup considered in~\cite{douglas}, where the orbifold action is along the transverse direction of the D-brane worldvolume rather than the longitudinal direction.} of the D3 branes $\mathbb{C}^2$ to make it $\mathbb{C}^2/\Gamma$. Since M-theory on $T^2$ is dual to IIB string theory, we lift the D3 branes to M5 branes by first taking the T-duality in the 4 direction to turn them into D4 branes and then blowing up the M-theory circle in the \# direction. The situation can be summarized in Table~\ref{table:2}.

\begin{table}[htbp]
\centering
\begin{tabular}{|c| ccccccccccc|}
\hline
\ & 0 &1 & 2 &  3 & 4  &  5 &  6 & 7 &  8 & 9 & \# \\ \hline
D3 & N & N& N& N& D & D & D&D&D& D & x \\ \hline
M5 & N & N& N& N& N & D & D&D&D&D&N \\ \hline
\end{tabular}
\caption{The string/M theory setup. Here, $D$ denotes ``Dirichlet'' and $N$ ``Neumann''. To avoid confusion, we use \# to deonte the 10th direction. Since the D3 brane exists only in the 9+1 dimensional universe, an ``x'' is put under the 10th direction to indicate that the D3 brane does not exist in that direction. \label{table:2}}
\end{table}
A stack of $N$ M5 branes whose worldvolume is placed on an ADE singularity $\mathbb{C}^2/\Gamma$ has a near horizon geometry $AdS_7/ \Gamma \times S^4$. Let the Lie algebra corresponding to the ADE singularity be $\mathfrak{g}(\Gamma)$. The boundary theory lives on directions 012345. Let the bulk direction be 6. We claim that there exists a coupling
\begin{equation}
\label{eq:duality1}
-\frac{1}{4\pi}\int \frac{C_3}{2\pi} \wedge \tr(F\wedge F)
\end{equation}
where $C_3$ is the 3-form in M-theory and $F$ is the 2-form field strength taking values in $\mathfrak{g}$. The integral is taken in the 456789\# directions. The 5-plane transverse to the M5 branes can be decomposed into $\mathbb{R}^+ \times S^4$, where $\mathbb{R}^+$ is the direction into the bulk, and $S^4$ is the 4-sphere that surrounds the M5 branes. Integrating by parts, we obtain
\begin{equation}
\frac{1}{4\pi} \int_{T^2 \times \mathbb{R}^+ \times S^4} \frac{dC_3}{2\pi } \wedge \tr(A\wedge dA + \frac{2}{3} A \wedge A\wedge A) = \frac{N}{4\pi} \int_{T^2 \times \mathbb{R}^+} \tr(A\wedge dA + \frac{2}{3} A \wedge A\wedge A)
\end{equation}
reproducing the Chern-Simons theory on $T^2$ at level $N$.

Before we derive this, we mention two similar constructions in the literature. Ref~\cite{dijkgraaf} reproduces a part of this duality by considering intersecting D4-D6 branes in type IIA string theory, so only the A-singularity is probed since D6 branes cannot create the D- or the E-singularity. Ref~\cite{vafawitten}, on the other hand, considers Nakajima's instanton construction~\cite{nakajima} on $\mathbb{R}^4/\Gamma$ using a certain twisted super Yang-Mills theory without explicitly mentioning the Chern-Simons theory. In~\cite{vafawitten}, it is shown that a certain combination of the middle-dimensional cohomology of $U(q)$ n-instanton moduli space on $\mathbb{R}^4/\Gamma$ has the same structure as the highest weight representation of a level $q$ state of algebra $\mathfrak{g}(\Gamma)$ (see equation 4.42 in~\cite{vafawitten}). The link to the Chern-Simons theory is implicit in their work because of the equivalence between WZW conformal blocks and Chern-Simons theory as shown in~\cite{wittenjones}. We derive the duality using a perspective different from these two constructions. Compared to~\cite{dijkgraaf}, our construction has the virtue of incorporating the D- and the E-singularity. Compared to~\cite{vafawitten}, our construction does not require twisting the super Yang-Mills theory and is easier to understand from a physicist's perspective.

The strategy we use to derive equation~\eqref{eq:duality1} is through the string-string duality. In the first step, we compactify the 4 direction and take the T-dual to end up in the Type IIA string theory. We now have $N$ D4 branes along the 01234 directions. The ADE singularity acting on the 0123 directions can be locally thought of as a certain degeneration limit of a K3 surface. Although the latter is compact and the former is noncompact, the distinction will not be important in the following derivation. It is known that heterotic string on $T^4$ is dual to IIA string on K3, with the coupling constants being related by the equation
\begin{equation}
    e^\phi = e^{-\phi'}
\end{equation}
where unprimed quantities are for the heterotic theory and the primed quantities are for the IIA theory~\cite{wittendynamics}. In addition, the NS-NS 2-forms are related in 6D as
\begin{equation}
dB = e^{-2\phi'} \star dB'
\end{equation}

Now, both the $T^4$ and the K3 are along the Euclideanized 0123 directions. Recall that the NS-NS 2-form in IIA theory descends from the 3-form in M-theory as $dB' \wedge dx^\# = dC_3$. So if one blows up the M-theory circle and lift the above relation to 7D (i.e. adding the $\#$ direction), one gets
\begin{equation}
dB = e^{-2\phi'} \star dC_3
\end{equation}

We now have $N$ M5 branes along the 01234\# directions, where the 4 and the \# directions are circles.

The heterotic theory has a modified action for its 3-form field strength (setting the curvature contribution to zero\footnote{It is amusing to note that if one were to include curvature, i.e. the gravitational Chern-Simons term, one would be able to give a string-theory derivation of the curvature counter term in the one-loop correction to the Chern-Simons action. In~\cite{wittenjones}, this counter term arises out of the consideration for canceling the framing anomaly.}):
\begin{equation}
\frac{1}{2}\int_{\mathbb{R}^6} e^{-2\phi} (dB -CS, dB-CS) 
\end{equation}
where $CS$ denotes the Chern-Simons three form and where we considered the $T^4$ compactified action so the integration is along the directions 456789 in the IIA theory. Notice that $e^{-2\phi} = e^{2\phi'}$ is the heterotic dilaton coupling. If we consider blowing up the M-theory circle to lift the above action to 7D, and plug in the duality relation, the above term becomes
\begin{equation}
\frac{1}{2}\int_{\mathbb{R}^7=T^2\times \mathbb{R}^5} e^{2\phi'} ( e^{-2\phi'} \star dC_3 -CS, e^{-2\phi'} \star dC_3 -CS) 
\end{equation}
where the integral is taken along the directions 456789\# and we split the directions along the torus (4\#) and the $\mathbb{R}^5$ (56789) transverse to the $N$ M5 branes (see Table 2). The $\mathbb{R}^5$ can be further decomposed into a radial direction times an $S^4$ that surrounds the $N$ M5 branes. From this expression, we get a cross term
\begin{equation}
-\int (CS, \star dC_3) = \int CS \wedge \star\star dC_3 = \int CS \wedge dC_3 = N \int CS
\end{equation}
where in the last step we integrated the 4-form fluxes along the $S^4$ to pick up the overall prefactor $N$, reproducing~\eqref{eq:duality1} and the Chern-Simons action we sought for.

In the IR limit, the Chern-Simons action is the unique action that has the lowest number of derivatives, and will characterize the ground state structure of the theory. Therefore, the ground state Hilbert space of $\mathcal{N}=4$ $U(q)$ SYM on $S^3/\Gamma$ is equivalent to that of the holographic dual which is the Hilbert space of level $q$ Chern-Simons theory with gauge group $\mathfrak{g}(\Gamma)$. This Chern-Simons Hilbert space is given by the set~\eqref{eq:a2}. In this work, we considered $\mathcal{N}=4$ $SU(q)$ SYM on $S^3/\Gamma$. Changing the gauge group from $U(q)$ to $SU(q)$ reduces the number of ground states in the SYM ground state Hilbert space. What is the corresponding reduction on the Chern-Simons side? To see this, we use the example $\Gamma=\mathbb{Z}_k$. The $SU(q)$ SYM ground states Hilbert space admits a level-rank duality: the dimension of $SU(q)$  ground state Hilbert on $S^3/\mathbb{Z}_k$ is the same as that of the $SU(k)$ ground state Hilbert space on $S^3/\mathbb{Z}_q$\footnote{A quick way to show this is to use the $SU(N)$ Ehrhart polynomials computed in~\eqref{eq:zkpoly2} and show that the $q$th term of $\text{Ehr}_{\mathfrak{su}(k)}$ is the same as the $k$th term of $\text{Ehr}_{\mathfrak{su}(q)}$.}. The only nontrivial subspace of the Chern-Simons Hilbert space that is invariant under the level-rank duality is the one given by the set~\eqref{eq:a3}~\cite{kac}. This concludes the physical derivation of~\eqref{eq:e1}.

\section{Conclusion}
\label{sec:9}
We have shown that the counting of root lattice states of level $q$ Chern-Simons theory Hilbert space on $T^2$ can be solved by computing the exact generating function using either the $\Omega$ operator calculus or the duality proposed in \cite{ganorandju} that has a similar origin to!\cite{dijkgraaf}\cite{vafawitten}. The reader might wonder why we framed the problem in terms of the Ehrhart polynomials. First, the number of root lattice states grows as some quasi-periodic polynomial, a property shared by Ehrhart polynomials. As we saw in section~\ref{sec:3}, the geometric formulation of the counting problem leads very naturally to the idea of Ehrhart polynomials. Second, and most importantly, recent development of mathematics and physics shows that it is always fruitful to find connections between different subfields of mathematics and physics. Since Ehrhart polynomials connect various branches of mathematics such as number theory and topology, formulating the problem using Ehrhart polynomials is an attempt at achieving more unification. \par

Despite being a decades-old subject, Chern-Simons theory is at the heart of the inter-connectedness explored in this paper. In fact, as shown in the last section and in ~\cite{ganorandju}\cite{dijkgraaf}\cite{vafawitten}, the Chern-Simons theory arise naturally at the long distance limit of some holographic system in string theory. Another key ingredient is the McKay correspondence~\cite{mckay}.  In fact, we saw in section~\ref{sec:7} that the simple Lie algebras know a lot more than the content of the McKay correspondence through the inverse of the Cartan matrices. The simple Lie algebras secretly know the determinant of the irreducible representations of the corresponding ADE subgroup. 

Finally, we pose some open questions. In this work, we focused on a special subspace of the Chern-Simons Hilbert space on $T^2$ given by the set~\eqref{eq:a3} and found that there is a description using a certain Supersymmetric Yang-Mills theory. Are there other Chern-Simons states that admit a similar dual description? Chern-Simons theory can be quantized on higher-genus Riemann surfaces. On the SYM side, compactifying the M5 branes on higher-genus Riemann surfaces will break another half of the supersymmetry. The building blocks of the Hilbert space on closed Riemann surfaces are the $N_{ijk}$ fusion coefficients. What does the SYM dual of the fusion coefficients look like? We will tentatively propose an answer to this question in a forthcoming work. In this work, we focused on the $SU(q)$ gauge group for the SYM theory, but one can also ponder over other compact Lie groups such as $SO(q)$ or $Sp(2q)$. What special states in the Chern-Simons theory would these representations correspond to? A standard way to construct $SO(q)$ and $Sp(2q)$ gauge groups in string theory is by using orientifolds, so we would expect that a similar brane construction using orientifolds can help shed light on this question.

\acknowledgments

We would like to thank Emil Albrychiewicz, Ori Ganor, Yasunori Nomura, and Tong Zhou for useful discussions. In particular, we thank Ori Ganor for pointing out the potential significance of the states considered in this paper.

%\pagebreak
%\bibliographystyle{my-h-elsevier}

%Author,
%\emph{Title},
%\emph{J. Abbrev.} {\bf vol} (year) pg.

\end{document}